\documentclass[12pt,preprint]{emulateapj}

\newcommand{\gray}{$\gamma$-ray}
\newcommand{\grays}{$\gamma$-rays}

\newcommand{\pubjournal}[5]{#4, #1, #2, #3}

\shorttitle{}
\shortauthors{Porter et al.}

\begin{document}

\title{Inverse Compton Emission from Galactic Supernova Remnants: 
Effect of the Interstellar Radiation Field}

\author{Troy A. Porter\altaffilmark{1}}
\affil{
  Santa Cruz Institute for Particle Physics,
  University of California, Santa Cruz, CA 95064
\email{tporter@scipp.ucsc.edu}}
\altaffiltext{1}{Also Department of Physics and Astronomy, 
Louisiana State University, Baton Rouge, LA 70803}

\author{Igor V. Moskalenko\altaffilmark{2}}
\affil{
   Hansen Experimental Physics Laboratory, 
   Stanford University, Stanford, CA 94305
\email{imos@stanford.edu}}
\altaffiltext{2}{Also Kavli Institute for Particle Astrophysics and Cosmology,
Stanford University, Stanford, CA 94309}

\and

\author{Andrew W. Strong}
\affil{Max-Planck-Institut f\"ur extraterrestrische Physik,
Postfach 1312, D-85741 Garching, Germany
\email{aws@mpe.mpg.de}}

\begin{abstract}


The evidence for particle
acceleration in supernova shells comes from electrons whose synchrotron
emission is observed in radio and X-rays. 
Recent observations by the HESS instrument
reveal that supernova remnants also emit TeV \grays; 
long awaited experimental evidence that supernova remnants can accelerate 
cosmic rays up to the ``knee'' energies. 
Still, uncertainty exists whether these \grays\ are produced by 
electrons via inverse Compton scattering or by protons via $\pi^0$-decay.
The multi-wavelength spectra of supernova remnants can be
fitted with both mechanisms, although a preference is often given to
$\pi^0$-decay due to the spectral shape at very high energies.
A recent study of the interstellar radiation field indicates
that its energy density, especially in the inner Galaxy, is higher
than previously thought.
In this paper we evaluate the effect of the interstellar radiation field
on the inverse Compton emission of electrons accelerated in a supernova remnant
located at different distances from the Galactic Centre.
We show that contribution of optical and infra-red photons to the inverse 
Compton emission
may exceed the contribution of cosmic microwave background and in some
cases broaden the resulted \gray\ spectrum. 
Additionally, we show that if a supernova remnant
is located close to the Galactic Centre
its \gray\ spectrum will exhibit a ``universal'' cutoff at very high
energies due to the Klein-Nishina effect and not due to the cut-off of the
electron spectrum.
As an example, we apply our calculations to the supernova remnants
RX J1713.7-3946 and G0.9+0.1 recently observed by HESS.
\end{abstract}


\section{Introduction}


A new calculation \citep{PS05} of the Galactic interstellar radiation 
field (ISRF)
consistent with multi-wavelength observations by DIRBE and FIRAS 
indicates that the energy density of the
ISRF is higher, particularly in the inner Galaxy, than previously 
thought. 
This has implications for the inverse Compton (IC) scattering 
of electrons and positrons and other electromagnetic 
processes in the interstellar medium \citep{Moskalenko2000,Moskalenko2006}.
Another place where the enhanced ISRF may play a role is the IC emission
off very high energy (VHE) electrons accelerated in a supernova 
remnant (SNR) shock.

SNRs are believed to be the primary sources of cosmic rays in the Galaxy.
Observations of X-ray \citep{Koyama1995} and \gray\ emission 
\citep{Aharonian2005b,Aharonian2006}
from SNR shocks reveal the presence of energetic particles,
thus testifying to efficient acceleration processes.
Acceleration of particles in collisionless shocks is a matter of
intensive research in conjunction with the problem of cosmic ray origin 
\citep{Drury1983,Blandford1987,Jones1991}.
Current models include nonlinear effects
\citep[e.g.,][]{Berezhko2006} and treat particle
acceleration using hydrodynamic codes 
\citep[e.g.,][]{Ellison2005}.
The predicted spectrum of accelerated
particles has a power-law form in rigidity with index which may
slightly vary around $-2.0$. 
Conventional estimates put the maximum
reachable particle energy (protons) at or just below the knee energy;
for the case of electrons the maximum energy is lower due to the synchrotron
and IC energy losses \citep{Ellison2005}.
Young SNRs may be capable of particle acceleration up to $10^{17}$ eV
due to the effect of magnetic field amplification \citep{Bell2001}
around the shock and assuming Bohm diffusion and a Sedov expansion law.

The VHE \gray\ emission from shell-type SNRs has been modelled using 
leptonic (IC) and hadronic ($\pi^0$-decay) scenarios 
\citep[e.g.,][]{Baring2005}. 
The leptonic scenario
fits the broad-band spectrum of a SNR assuming a pool of accelerated
electrons scattered off the cosmic microwave background (CMB)
producing VHE \grays\ while the magnetic field and electron spectrum cut-off 
are tuned to fit the radio and X-ray data \citep[e.g.,][]{Lazendic2004}. 
The hadronic model fits the VHE \gray\
spectrum assuming a beam of accelerated protons hits a target, such as a 
nearby molecular cloud \citep{Aharonian2002}.
The latter, if definitively
proven, would be the first experimental evidence of proton acceleration
in SNRs.
While there is no clear distinction between different models of VHE
emission from SNRs, some authors tend to prefer the hadronic scenario
since it fits better the observed spectral shape 
\citep{Aharonian2006,Berezhko2006}.
Such a preference, however, is made based on a simplified ``one-zone''
model which typically includes CMB photons only.

In this paper we evaluate the effect of the ISRF on the IC emission of
electrons accelerated in SNRs located at different distances from
the Galactic Centre (GC). 
As examples, we apply our calculations to the shell-type SNR
RX J1713.7-3946 and composite SNR G0.9+0.1 recently observed by HESS 
\citep{Aharonian2005a, Aharonian2006}. 

\section{Interstellar Radiation Field}

The Galactic ISRF calculation uses a model for the 
distribution of stars in the Galaxy, a model for the dust distribution and 
properties, and a treatment of scattering, absorption, and subsequent 
reemission of the stellar light by the dust.
A brief description of our calculation is available in \citet{Moskalenko2006}; 
we re-iterate the main points here.

Our stellar model assumes a type classification based on that used in the SKY 
model of \cite{Wainscoat1992}, with modifications to account for 
results from recent experiments such as 2MASS, SDSS, and others 
\citep[e.g.,][]{Ojha2001,Juric2005}, and synthetic 
spectral modelling studies \citep[e.g.,][]{Girardi2002}.

Dust is modelled with a mixture of
polycyclic aromatic hydrocarbons, graphite, and silicate. 
The dust grains are spherical and we include details of their 
absorption and scattering 
efficiencies, abundances, and size distribution in the scattering and 
heating calculations \citep{Li2001,Weingartner2001}.
Stochastic and equilibrium heating of the dust grains is treated following 
\cite{Draine2001} and \cite{Li2001}.
The dust is assumed to follow the Galactic gas distribution and metallicity
gradient \citep[][and references therein]{Moskalenko2002,Strong2004b}.

\begin{figure}[t]
\centerline{\includegraphics[width=3.2in]{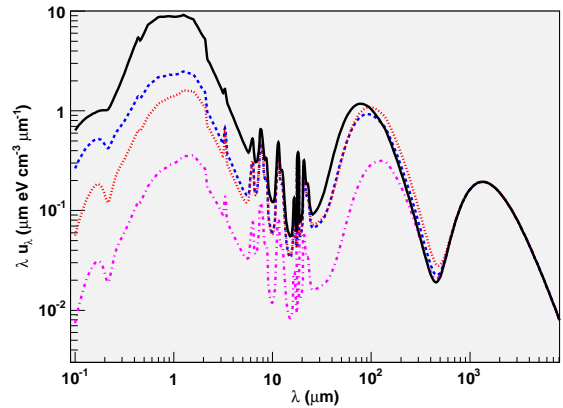}}
\caption{
Interstellar radiation field energy density:
solid line, $R = 0$ kpc, $z = 0$ kpc;
dashed line, $R = 3$ kpc, $z = -0.05$ kpc;
dotted line, $R = 4$ kpc, $z = 0$ kpc;
dash-dotted line, $R = 7.5$ kpc, $z = 0$ kpc.}
\label{fig1}
\end{figure}

Figure~\ref{fig1} shows the total ISRF at the positions 
we will calculate the IC emission in Section~\ref{SecCalc}.
Information on the calculation method, spatial and spectral 
distribution in the Galactic plane, and a brief comparison of our 
calculations with previous results and the locally observed ISRF are 
given in \cite{Moskalenko2006}; 
further discussion of the new ISRF is deferred to a forthcoming paper (Porter
\& Strong, in preparation).

\section{Calculations}
\label{SecCalc}

For an assumed isotropic ISRF and electron distribution, 
the IC emissivity (photons cm$^{-3}$ s$^{-1}$) 
is given by 

\begin{equation}
\frac{dP_\gamma}{d\epsilon_2} = 
c \int d\epsilon_1 \int d\gamma_e \,
n_\epsilon(\epsilon_1) n_e(\gamma_e) 
\frac{dQ_\gamma(\gamma_e, \epsilon_1)}{d\epsilon_2},
\label{ICSEmissivity}
\end{equation}

\noindent
where 
$\epsilon_2 = E_\gamma/m_e c^2$ and $\epsilon_1 = \epsilon/m_e c^2$
are the dimensionless \gray\ and target photon energy,
$\gamma_e = E_e /m_e c^2$ is the electron Lorentz factor, 
$n_\epsilon$ and $n_e$ are the
differential target photon and electron number densities, and 
$dQ_\gamma/d\epsilon_2$ is given by eq.(9) in \cite{Jones1968}.

To illustrate the effect of different electron spectral indices on the
IC emission, we take an electron number density
$n_e (\gamma_e) = n_0 \gamma_e^{-\delta}$ 
normalised so that $\int_{\gamma_{\rm min}} ^\infty n_e (\gamma_e) d\gamma_e = 1$ 
cm$^{-3}$ with $\gamma_{\rm min} = 1 \, {\rm MeV}/m_e c^2$  for 
$\delta = 1.8$, $\delta = 2.0$ and $\delta = 2.5$.

\begin{figure}[h]
\centerline{\includegraphics[width=3.2in]{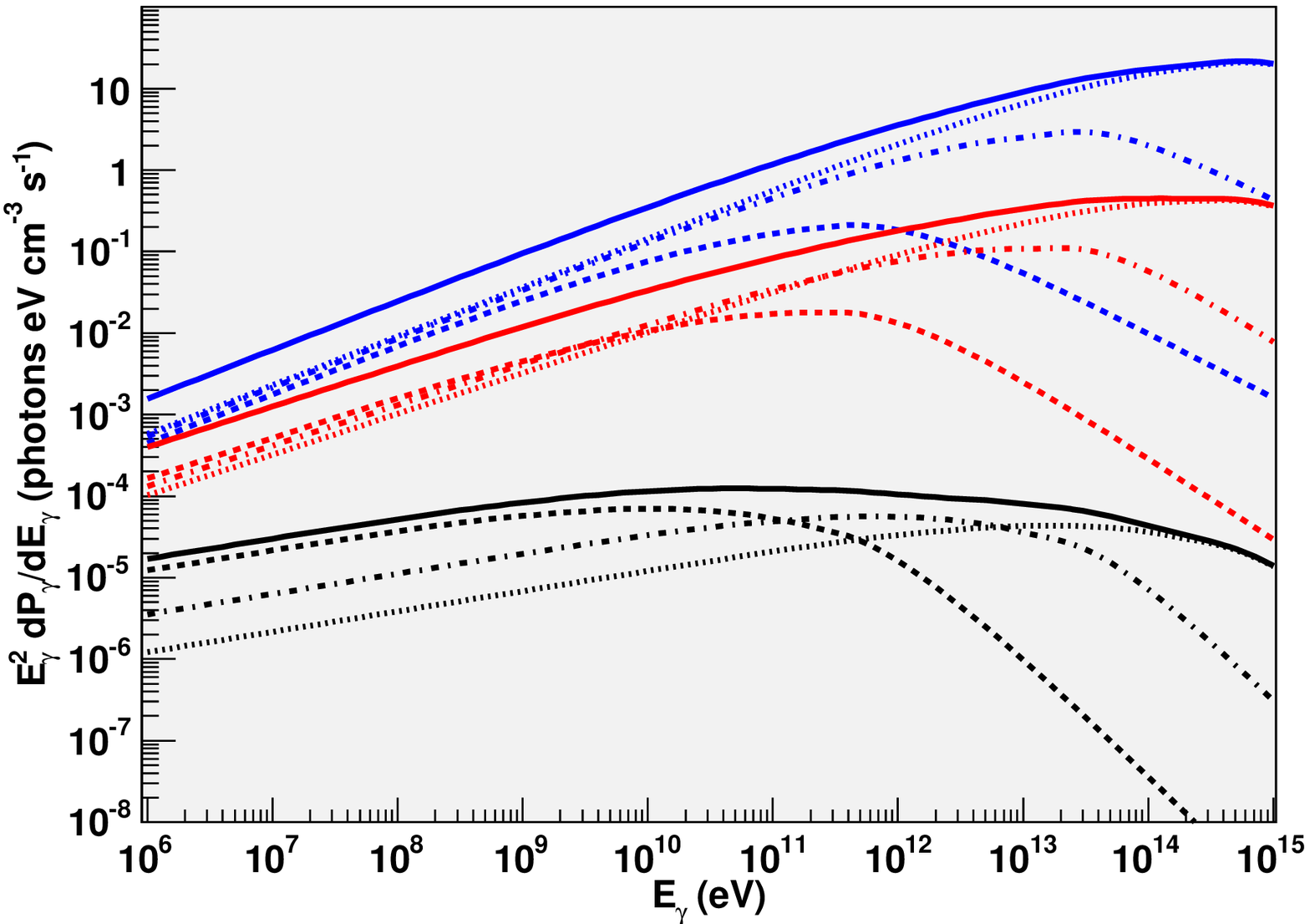}}
\centerline{\includegraphics[width=3.2in]{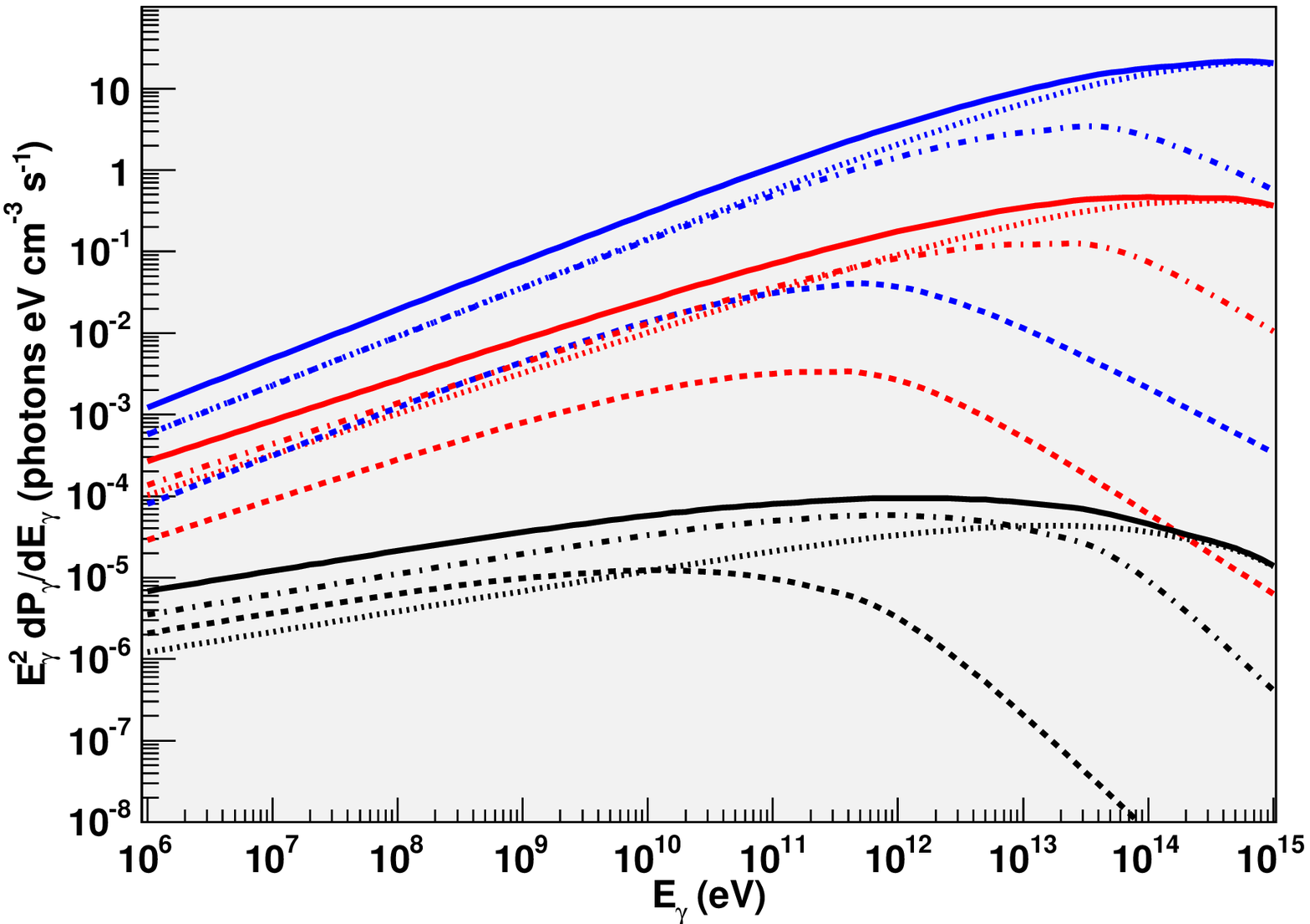}}
\caption{Inverse Compton emissivity in the Galactic plane for $R = 0$ kpc 
(top) and $R = 4$ kpc (bottom).
Line styles: solid line, total; dashed, optical; dot-dashed, infra-red; 
dotted, CMB.
For each panel, top-most line-sets correspond to $\gamma_e^{-1.8}$, middle
line-sets correspond to $\gamma_e^{-2.0}$ and bottom-most
correspond to $\gamma_e^{-2.5}$ electron spectra respectively.}
\label{fig2}
\end{figure}

Figure~\ref{fig2} shows the calculated IC emissivity using
the ISRF at $R = 0$ (upper) and $R = 4$ (lower) in the Galactic plane.
There are two essential effects that contribute to produce the total emission
curves in Fig.~\ref{fig2}.
First, the variation in the electron spectral index 
$\delta = 1.8 \rightarrow 2.5$ 
increases the contribution by optical and 
infra-red (IR) 
photons to the \gray\ emission below $\sim 50$ GeV and $\sim 1$ TeV,
respectively.  
As $\delta$ increases, the number of low energy electrons increases;
the IC scattering rate by these electrons off optical and
IR photons becomes larger giving more upscattered photons.
In turn, as the \gray\ energy increases, the contribution by the optical
and IR photons
to the total emissivity decreases from the reduction of the scattering
cross-section in the Klein-Nishina (KN) regime.
Thus, softer electron spectra promote more efficient scattering of the 
optical and IR components, increasing their relative contributions.
Second, 
the density of optical and IR photons varies with position in the
Galaxy.
Toward the GC, the optical photon density is high, 
almost an order of magnitude larger per target photon energy 
interval when compared with the optical component at $R = 4$ kpc.
Similarly, over the inner Galaxy the IR photon energy density is 
significantly higher than the CMB.
This gives a contribution by the optical and IR photons to the 
total IC emission 
that is at least comparable to the CMB over these regions of the Galaxy.

Note that the marked decrease in the optical, then IR, 
then CMB emission as the \gray\ energy increases in Fig.~\ref{fig2} 
is due to the reduction of the scattering cross-section in the 
KN regime, as described above, and not due to 
any cut-off we have introduced in the electron spectrum; if a 
cut-off was included in the electron spectrum, it would further decrease
the contribution from the CMB relative to the optical and IR
components (see Fig.~\ref{fig4}).

To illustrate the effect of the ISRF in a purely 
leptonic scenario,
we calculate the synchrotron and IC emission using a one-zone model 
\citep{Aharonian1999} for the SNR RXJ1713.7-3946. 
We use an electron spectrum 
$Q_e(E) = Q_0 E^{-\delta} \exp \left( -E/E_{\rm max} \right)$
where $Q_0$, $\delta$ and $E_{\rm max}$ are adjustable parameters.
The synchrotron emission for a magnetic field strength $B$ 
is calculated using the usual formulae 
\citep{Ginzburg1979,Ghisellini1988}. 

\begin{figure}[!t]
\centerline{\includegraphics[width=3.2in]{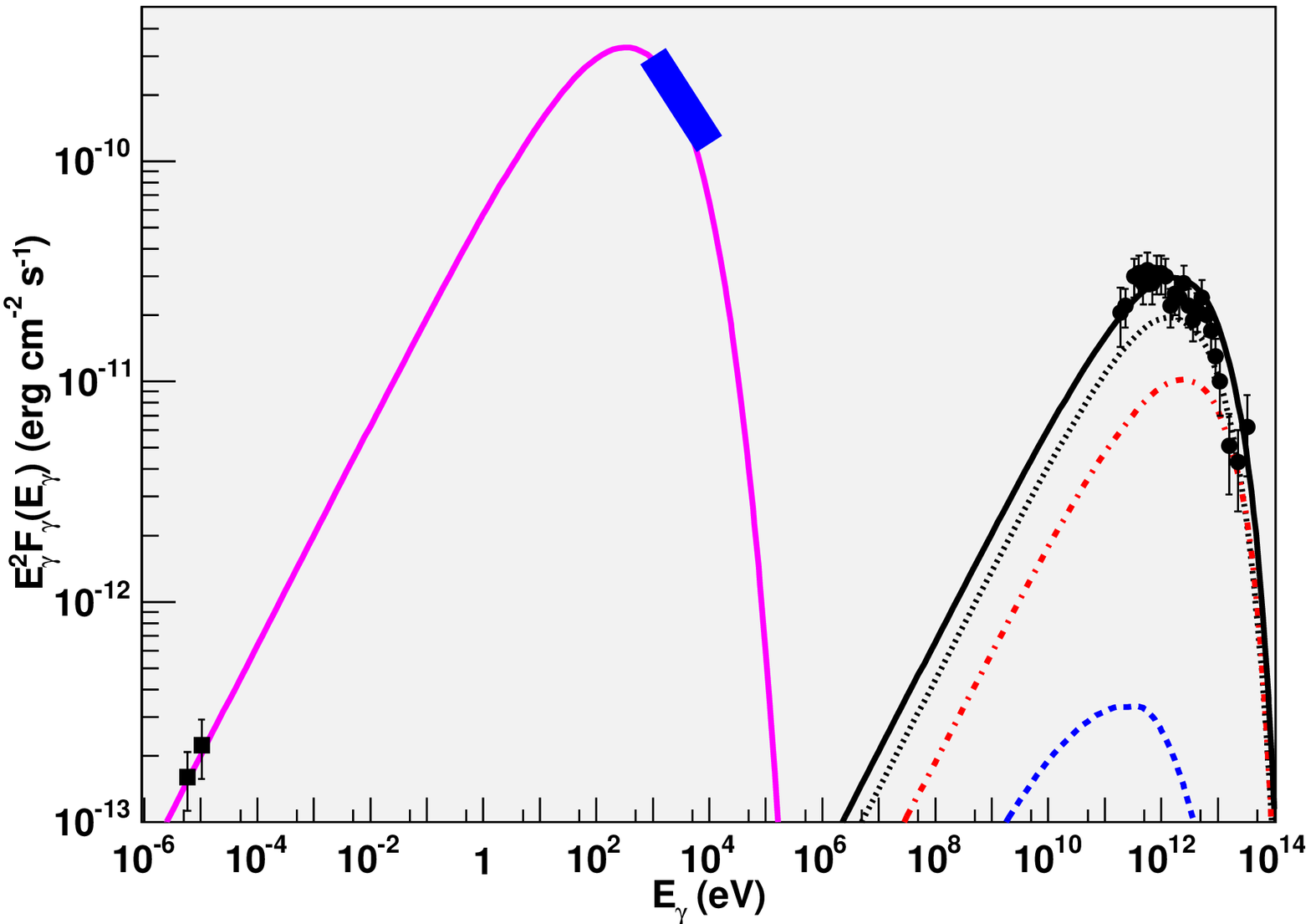}}
\centerline{\includegraphics[width=3.2in]{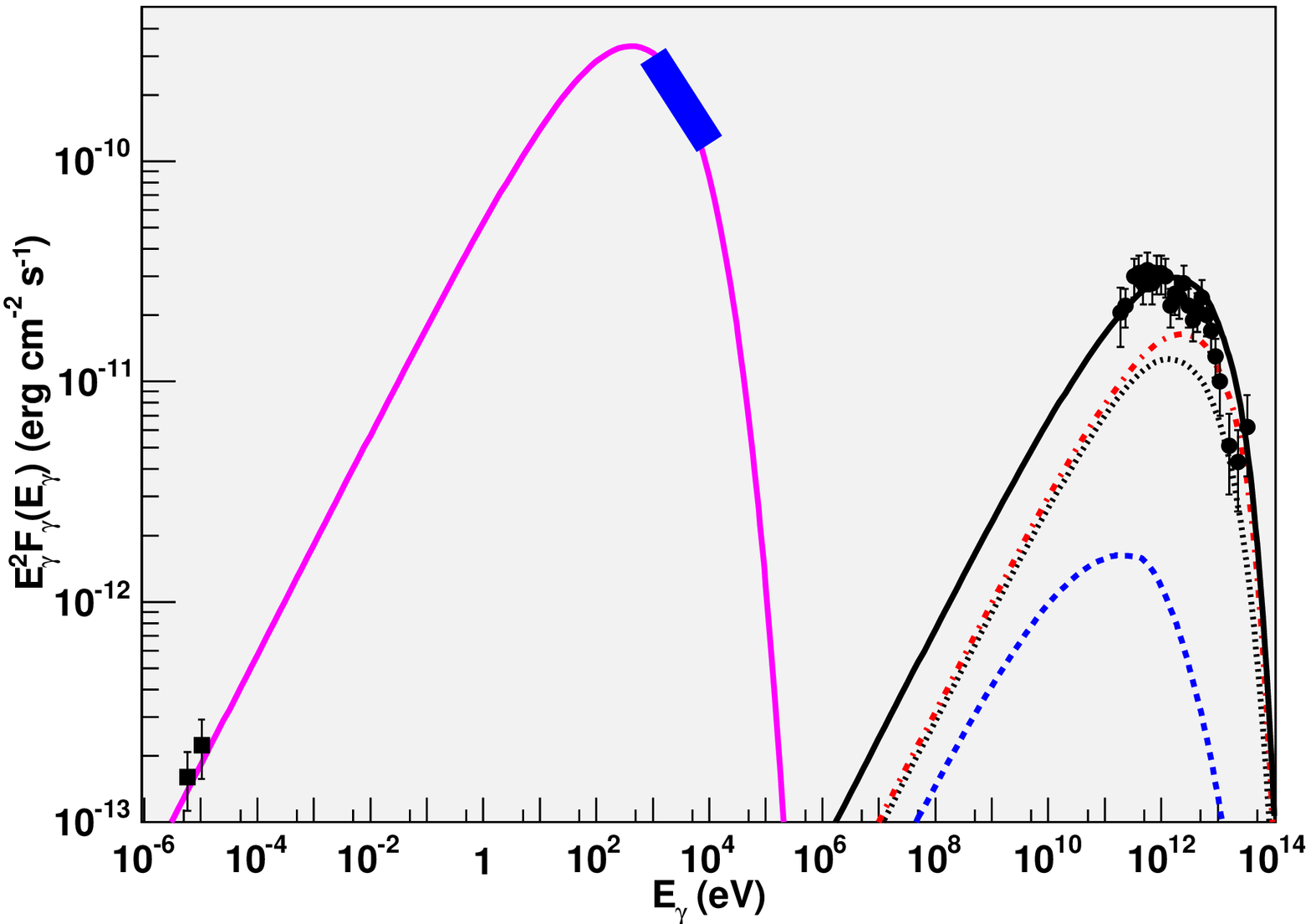}}
\caption{Fit to the flux spectrum of RXJ1713.7-3946 using the ISRF 
for $d = 1$ kpc (upper) and $d = 6$ kpc (lower).
Line-styles: solid, total synchrotron and IC flux; dashed, optical IC; 
dash-dotted, infra-red IC; dotted, CMB IC.
Data: radio (ATCA) from \citet{Lazendic2004} for the north-west rim
of RXJ1713.7-3946, but upscaled by a factor of 2 following 
\citet{Aharonian2006};
X-ray (ASCA) from \citet{Pannuti2003};
\gray\ (HESS) from \citet{Aharonian2006}.}
\label{fig3}
\end{figure}

\begin{figure}[h]
\centerline{\includegraphics[width=3.2in]{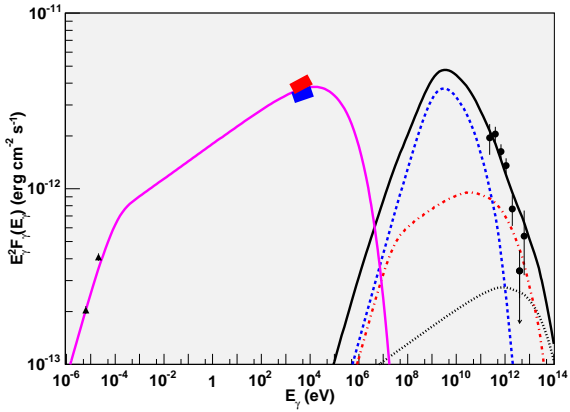}}
\caption{Fit to the flux spectrum of G0.9+0.1.
Line-styles as for Fig.~\ref{fig3}.
Data: radio (VLA) from \citet{Helfand1987};
X-ray (BeppoSAX) at 1.8--10 keV from \citet{Sidoli2000} (upper solid region) 
and (XMM) at 2--12 keV from \citet{Porquet2003} (lower solid region);
\gray\ (HESS) from \citet{Aharonian2005a}.}
\label{fig4}
\end{figure}

Figure~\ref{fig3} shows the photon flux from RXJ1713.7-3946 calculated 
for a distance of $d = 1$ kpc (upper) and $d = 6$ kpc (lower).
This covers the range of distances quoted by various authors 
\citep{Lazendic2004,Aharonian2006}.
The IC emission is calculated using the appropriate ISRF spectrum
for each distance 
(see Fig.~\ref{fig1}).
For both cases, we reproduce the TeV emission first: with  
$\delta \sim 1.8-2.2$ and $E_{\rm max} \sim 15-40$ TeV a reasonable fit
can be obtained.
Next, we fit the radio and X-ray data within the constraints
imposed by the TeV emission: $\delta \sim 2.0$ and $E_{\rm max} \sim 15-25$ TeV, 
with $B \sim 5-20$ $\mu$G are allowed.
The electron luminosity for an age
of 1000 years ($d = 1$ kpc) or 10000 years ($d = 6$ kpc) is 
$(1-4) \times 10^{37}$ erg s$^{-1}$ with
larger luminosities for the more distant case.
The parameters used 
are not unique.
As a particular example, the values used in 
Fig.~\ref{fig3} are: 
$\delta = 2.0$, electron luminosity of $1.5 \times 10^{37}$ erg s$^{-1}$, 
$E_{\rm max} = 25$ TeV, and $B = 12$ $\mu$G ($d = 1$ kpc);
$\delta = 2.0$, 
electron luminosity of $3.5 \times 10^{37}$ erg s$^{-1}$, $E_{\rm max} = 25$ TeV,
and $B = 15$ $\mu$G ($d = 6$ kpc).

We also consider the composite SNR G0.9+0.1 observed by HESS 
\citep{Aharonian2005b};
this composite SNR is a source of VHE electrons, as evidenced by the 
synchrotron X-ray emission \citep{Porquet2003}, and is located in the GC 
region where the ISRF density is high.
As for RXJ1713.7-3946, we use a one-zone model except we modify the source 
spectrum to be a broken power-law with a break at energy $E_{\rm break}$.
As before, we adjust the electron luminosity and 
high energy spectral index to fit the TeV 
observations, then vary $B$ and $E_{\rm max}$ to reproduce the X-ray data.
Then, $E_{\rm break}$ and the low energy spectral index are adjusted to 
give consistency with the radio data.
Figure~\ref{fig4} shows the calculated intrinsic 
photon flux for G0.9+0.1 at the GC.
The parameter values are: $\delta_L = 2.0$, 
$\delta_U = 2.8$, $E_{\rm break} = 15$ GeV, 
where $\delta_L$ and $\delta_U$ are the spectral indices
below and above $E_{\rm break}$ respectively, $B = 9.5$ $\mu$G, 
$E_{\rm max} = 1000$ TeV, and an electron luminosity 
$2.5\times10^{38}$ erg s$^{-1}$ for an assumed age of 5000 years.
Note that these values are again not unique, and other combinations are
possible.

The cut-off in the IC emission for the optical 
and IR in Figs.~\ref{fig3} and~\ref{fig4} 
comes from the reduction of the scattering cross-section
in the KN regime; only for the CMB is the IC spectral cut-off
solely due to the cut-off in the electron spectrum.
The cut-off in the IC emission for the IR in Fig.~\ref{fig3} is only
partially due to the KN effect since the electron spectral 
cut-off energy is essentially lower.
This effect plays an important role for a SNR close to the GC where the
ISRF is dominant.
In this case, the cut-off energy of the VHE \gray\ spectrum becomes 
universal and 
defined exclusively by the reduction of the scattering cross-section and not
by an assumed electron spectral cut-off.
Note also, 
for energies higher than $\sim 10$ TeV, the intrinsic spectrum will be 
further reduced by attenuation due to pair production 
on the ISRF itself \citep{Moskalenko2006}.
  
\section{Discussion}

We have evaluated the effect of the ISRF on the IC emission of electrons 
accelerated in SNRs. 
Our calculation shows that for a SNR located in the inner Galaxy
the contribution of optical and IR photons to the IC emission
exceeds the contribution of the CMB and in some
cases broadens the resultant \gray\ spectrum.
SNRs located close to the GC exhibit a ``universal'' cut-off
at VHEs due to the KN effect and not due to a cut-off in the
electron spectrum.

We have made a calculation of the IC and synchrotron
emission for simple power-law
electron spectra including the contribution by the ISRF
using the shell-type SNR RXJ1713.7-3946 and composite SNR G0.9+0.1 as examples.
Within the confines of a simple one-zone leptonic model, it is possible to 
fit the observed flux spectrum with a reasonable combination of model 
parameters.
Observations in the GeV to sub-TeV range by the GLAST experiment will be 
critical in distinguishing between leptonic or hadronic scenarios of \gray\
production in SNRs as the predictions for the spectral shape in this energy 
range are distinctly different.

Finally, we point out at least two cases for the leptonic model 
where the contribution of the 
ISRF will be critical.
First will be when fitting the synchrotron peak dictates 
a relatively {\it low-energy} cut-off in the electron spectrum.
In this case, the VHE \grays\ will be produced by lower-energy electrons
scattered off optical and IR photons. 
Second will be when fitting the synchrotron peak dictates 
a {\it high-energy} cut-off in the electron spectrum.
An equal or exceeding contribution of optical and IR protons, 
but with lower energy cut-off
in the spectrum of upscattered \grays\ due to the KN effect,
will effectively broaden the IC peak.
Observations of the TeV emission from SNRs in the inner Galaxy may thus 
serve as a probe of the ISRF near their location.
In turn, observations of shell-type SNRs in the outer Galaxy where the CMB 
photons will provide
the majority of the IC emission may be used to evaluate the spectrum of
accelerated electrons in the shell and for studies of shock acceleration.

\acknowledgments
We thank Olaf Reimer for useful discussions.
T.\ A.\ P.\ acknowledges partial support from the US Department of Energy.
I.\ V.\ M.\ acknowledges partial support from NASA
Astronomy and Physics Research and Analysis Program (APRA) grant.

\clearpage


\begin{thebibliography}{}

\bibitem[Aharonian \& Atoyan(1999)]{Aharonian1999}
  Aharonian, F. \& Atoyan, A. M., \pubjournal{\aap}{351}{330}{1999}{}

\bibitem[Aharonian(2002)]{Aharonian2002}
  Aharonian, F., \pubjournal{Nature}{416}{797}{2002}{}

\bibitem[Aharonian et al.(2005a)]{Aharonian2005a}
  Aharonian, F., et al., \pubjournal{\aap}{432}{L25}{2005a}{}

\bibitem[Aharonian et al.(2005b)]{Aharonian2005b}
  Aharonian, F., et al., \pubjournal{\aap}{437}{L7}{2005b}{}

\bibitem[Aharonian et al.(2006)]{Aharonian2006}
  Aharonian, F., et al., \pubjournal{\aap}{449}{223}{2006}{}



\bibitem[Baring et al.(2005)]{Baring2005}
  Baring, M. G., Ellison, D. C., \& Slane, P. O., 
  \pubjournal{Adv. Sp. Res.}{35}{1041}{2005}{}

\bibitem[Bell \& Lucek(2001)]{Bell2001}
  Bell, A. R. \& Lucek, S. G., \pubjournal{\mnras}{321}{433}{2001}{}

\bibitem[Berezhko \& V\"{o}lk(2006)]{Berezhko2006}
  Berezhko, E. G. \& V\"{o}lk, H. J., \pubjournal{\aap}{451}{981}{2006}{}

\bibitem[Blandford \& Eichler(1987)]{Blandford1987}
  Blandford, R. \& Eichler, D., \pubjournal{Phys. Rep.}{154}{18}{1987}{}

\bibitem[Draine \& Li(2001)]{Draine2001}
  Draine, B. T. \& Li, A., \pubjournal{\apj}{551}{807}{2001}{}

\bibitem[Drury(1983)]{Drury1983}
  Drury, L., \pubjournal{Sp. Sci. Rev.}{36}{57}{1983}{}
  
\bibitem[Ellison \& Cassam-Chena\"{i}(2005)]{Ellison2005}
  Ellison, D. C. \& Cassam-Chena\"{i}, G., \pubjournal{\apj}{632}{920}{2005}{}



\bibitem[Ghisellini et al.(1988)]{Ghisellini1988}
  Ghisellini, G., Guilbert, P. W., \& Svensson, R., 
  \pubjournal{\apj}{334}{L5}{1988}{}

\bibitem[Ginzburg(1979)]{Ginzburg1979}
  Ginzburg, V. L., ``Theoretical physics and astrophysics'', 
  International Series in Natural Philosophy (Oxford: Pergamon) (1979)

\bibitem[Girardi et al.(2002)]{Girardi2002}
  Girardi, L., et al., \pubjournal{\aap}{391}{195}{2002}{}


\bibitem[Helfand \& Becker(1987)]{Helfand1987}
  Helfand, D. J. \& Becker, R. H., \pubjournal{\apj}{314}{203}{1987}{}


\bibitem[Henyey \& Greenstein(1941)]{Henyey1941}
  Henyey, L. G. \& Greenstein, J. L., \pubjournal{\apj}{93}{70}{1941}{}

\bibitem[Jones(1968)]{Jones1968}
  Jones, F. C., \pubjournal{Phys. Rev.}{167}{1159}{1968}{}

\bibitem[Jones \& Ellison(1991)]{Jones1991}
  Jones, F. C. \& Ellison, D. C., \pubjournal{Sp. Sci. Rev.}{58}{259}{1991}{}

\bibitem[Juric et al.(2005)]{Juric2005}
  Juric, M., et al., ApJ submitted, 2005, arXiv : (astro-ph/0510520)

\bibitem[Koyama et al.(1995)]{Koyama1995}
  Koyama, K., et al., \pubjournal{Nature}{378}{255}{1995}{}

\bibitem[Lazendic et al.(2004)]{Lazendic2004}
  Lazendic, J. S., et al., \pubjournal{\apj}{602}{271}{2004}{}

\bibitem[Li \& Draine(2001)]{Li2001}
  Li, A. \& Draine, B. T., \pubjournal{\apj}{554}{778}{2001}{}


\bibitem[Mereghetti et al.(1998)]{Mereghetti1998}
  Mereghetti, et al., \pubjournal{\aap}{331}{L77}{1998}{}

\bibitem[Moskalenko \& Strong (2000)]{Moskalenko2000}
  Moskalenko, I. V. \& Strong, A. W., \pubjournal{\apj}{528}{357}{2000}{}

\bibitem[Moskalenko et al.(2002)]{Moskalenko2002}
  Moskalenko, I. V., Strong, A. W., Ormes, J. F., \& Potgeiter, M. S.,
  \pubjournal{\apj}{565}{280}{2002}{}

\bibitem[Moskalenko et al.(2006)]{Moskalenko2006}
  Moskalenko, I. V., Porter, T. A., \& Strong, A. W., 
  \pubjournal{\apj}{640}{L155}{2006}{}

\bibitem[Ojha(2001)]{Ojha2001}
  Ojha, D. K., \pubjournal{\mnras}{322}{426}{2001}{}

\bibitem[Pannuti et al.(2003)]{Pannuti2003}
  Pannuti, T. G., et al. \pubjournal{\apj}{593}{377}{2003}{}

\bibitem[Porquet et al.(2003)]{Porquet2003}
  Porquet, D., et al., \pubjournal{\aap}{401}{197}{2003}

\bibitem[Porter \& Strong(2005)]{PS05}
  Porter, T. A. \& Strong, A. W., 2005,  
  in Proc.\ $29^{\rm th}$ Int.\ Cosmic Ray Conf.\ (Pune) (astro-ph/0507119)


\bibitem[Sidoli et al.(2000)]{Sidoli2000}
  Sidoli, L., et al., \pubjournal{\aap}{361}{719}{2000}{}


\bibitem[Strong et al.(2004)]{Strong2004b}
  Strong, A. W., Moskalenko, I. V., Reimer, O., Digel, S. \& Diehl, R., 
  \pubjournal{\aap}{422}{L47}{2004}{}

\bibitem[Wainscoat et al.(1992)]{Wainscoat1992}
  Wainscoat, R. J., et al., \pubjournal{\apjs}{83}{111}{1992}{}

\bibitem[Weingartner \& Draine(2001)]{Weingartner2001}
  Weingartner, J. C. \& Draine, B. T., \pubjournal{\apj}{548}{296}{2001}{}

\end{thebibliography}
\end{document}